\documentclass[12pt]{article}

\usepackage{amsfonts}
\usepackage{amssymb}
\usepackage{amsmath}
\usepackage{latexsym}

\def\beq{\begin{equation}}
\def\eeq{\end{equation}}

\def\bq{\begin{quote}}
\def\eq{\end{quote}}

\newcommand{\non}{\nonumber}
\newcommand{\be}{\begin{equation}}
\newcommand{\ee}{\end{equation}}
\newcommand{\bea}{\begin{eqnarray}}
\newcommand{\eea}{\end{eqnarray}}
\newcommand{\ba}{\begin{array}}
\newcommand{\ea}{\end{array}}
\newcommand{\al}{\alpha}

\newcommand{\pa}{\partial}
\newcommand{\ep}{\epsilon}
\newcommand{\si}{\sigma}
\newcommand{\la}{\lambda}

\newcommand{\ta}{\tau}

\newcommand{\om}{\omega}
\newcommand{\Om}{\Omega}

\newcommand{\rar}{\rightarrow}

\allowdisplaybreaks
\newcounter{mycount}

\begin{document}

\begin{titlepage}
\vspace{-1mm}
\begin{flushright}
Preprint ICN-UNAM 03-08\\
July 14, 2003
\end{flushright}
\vspace{1mm}
\begin{center}{\bf\Large\sf Perturbations of integrable systems and
Dyson-Mehta integrals}
\end{center}

\begin{center}

{\bf Alexander V.~Turbiner{\normalsize \footnote{
turbiner@nuclecu.unam.mx}${}^{,} $\footnote{On leave of absence
from the Institute for Theoretical and Experimental Physics, \\
\indent \hspace{5pt} Moscow 117259, Russia.}}}\\ {\em Instituto de
Ciencias Nucleares, UNAM, A.P. 70-543, 04510 M\'exico}
\\[2mm]
{\bf\large Abstract}
\end{center}
\small{
\begin{quote}
We show that the existence of algebraic forms of quantum,
exactly-solvable, completely-integrable $A-B-C-D$ and $G_2, F_4,
E_{6,7,8}$ Olshanetsky-Perelomov Hamiltonians allow to develop the
{\it algebraic} perturbation theory, where corrections are
computed by pure linear algebra means. A Lie-algebraic
classification of such perturbations is given. In particular, this
scheme admits an explicit study of anharmonic many-body problems.
The approach also allows to calculate the ratio of a certain
generalized Dyson-Mehta integrals algebraically, which are
interested by themselves.
\end{quote}
}
\vskip .5cm

\begin{center}
{\it Invited talk given at the Workshop "Superintegrable systems
in classical and quantum mechanics",\\ Montreal, Canada,
\\September, 2002} \\[10pt]
To be published in Proceedings
\end{center}

\end{titlepage}
\vskip 2mm

\setcounter{equation}{0}

\section{Introduction}

This talk is devoted to a study of perturbations of the
quantum-mechanical integrable systems. The feature of
integrability of these systems attract a lot of attention which in
our opinion is not fully justified. The main problem of quantum
mechanics is {\it the finding the spectra of the Hamiltonian via
solving the Schroedinger equation}. However, a knowledge of
operators commuting to the Hamiltonian does not help to solve the
main problem of quantum mechanics. At least, for the moment such a
connection is not known. The simplest illustration of the
situation is provided by the one-dimensional quantum dynamics - by
definition, any one-dimensional dynamics is completely- and
super-integrable - but it does not imply that a general
one-dimensional Schroedinger equation with arbitrary potential can
be solved exactly in whatever sense. The problem remains to be
transcendental and is equivalent to a diagonalization of a generic
matrix of infinite size. Existence of the integrals commuting to
the Hamiltonian gives a chance to assign definite quantum numbers
to the eigenstate, which also can be done {\it constructively} for
the only situation of knowledge of exact solutions of the original
Schroedinger equation. However, among known integrable systems in
quantum mechanics there exists a class of the systems with an
outstanding property -- these integrable systems are {\bf
exactly-solvable} in a sense that their eigenfunctions and
eigenvalues can be found exactly. The notion {\it exactly} needs
to be explained and we introduce a notion of {\it
exact-solvability}.

Let us take Hamiltonian
\[
{\cal H}\ =\ -\Delta \ +\ V(x)\ ,\qquad x \in R^d
\]
The system is called {\it integrable}, if there exist $k$
functionally-independent operators commuting with ${\cal H}$,
\[
      [{\cal H}, I_i]\ =\ 0\ , \qquad i=1,2\ldots k
\]
If $k > d-1$, the system is called {\it superintegrable}. If
$k=d-1$ and for all $i,j$
\[
     [I_i, I_j]=0
\]
the system is called {\it completely-integrable}.

The main problem of quantum mechanics is to solve the Schroedinger
equation
\[
      {\cal H} \Psi (x)\ =\ E \Psi (x) \quad , \quad \Psi (x) \in L^2
      (R^d)\ .
\]
Integrability in whatever sense does not help to solve the
Schroedinger equation. However, among integrable systems there are
several {\bf exactly-solvable} systems. They can be used as zero
approximations to study realistic physical systems.

Now we introduce a definition of {\it exact-solvability}. Let us
assume that a linear operator $h$ possesses infinitely-many
finite-dimensional invariant subspaces ${\cal V}_n,\quad
n=0,1\ldots$, which can ordered
\[
{\cal V}_0 \subset  {\cal V}_1 \subset {\cal V}_2 \subset \ldots
 \subset  {\cal V}_n  \subset \ldots {\cal V}\ ,
\]
thus forming an {\em infinite flag (filtration)} ${\cal V}$. Hence
the operator $h$ preserves the flag ${\cal V}$.

{\bf General Definition } \cite{Turbiner:1994}
\begin{quote}
An operator $h$ which preserves an infinite flag of
finite-dimensional spaces ${\cal V}$ is called {\it
exactly-solvable operator with flag ${\cal V}$}.

If given $h$ preserves several flags and among them there is a
flag for which $\mbox{dim} {\cal V}_n$ is {\it maximal} for any
given $n$, such a flag is called {\it minimal}.
\end{quote}

Below we deal with certain linear spaces of polynomials in several
variables. It leads to a notion {\it `characteristic vector'}
\cite{Ruehl:1998}. Let us consider the triangular linear space of
polynomials in $k$ variables
\begin{equation}
\label{flag}
 {\cal P}_{n}^{(\al_1, \ldots, \al_k)} \ = \ \langle s_1^{p_1}
s_2^{p_2} \ldots s_k^{p_k} | 0 \leq \al_1 p_1 + \al_2 p_2 +\ldots
+ \al_k p_k \leq n \rangle\ ,
\end{equation}
where $\al$'s are positive numbers and $n=0,1,2,\ldots$.
Characteristic vector is a vector with components which are equal
to the coefficients (weights) $\al_i$ in front of $p_i$:
\begin{equation}
\label{char.vec}
 \vec f = (\al_1, \al_2, \ldots \al_k)\ .
\end{equation}
Taking the spaces with $n=0,1,2,\ldots$ we arrive at the flag
which has ${\cal P}_{n}^{(\al_1, \ldots, \al_k)}$ as {\it
generating} linear space. We call this flag ${\cal P}^{(\al_1,
\ldots, \al_k)}$. The flag with $\vec f_0 = (1,1,\ldots,1)$ is
called {\it basic} and denoted ${\cal P}^{(k)} \equiv {\cal
P}^{(1, \ldots, 1)}$. It is clear that is minimal among possible
flags ever. This flag has a Lie-algebraic interpretation.

Let us take the algebra $gl_{k+1}$ in almost degenerate or totally
symmetric representation characterized by spins
$(n,\underbrace{0,0,\ldots 0}_{k-1})$,
\begin{eqnarray}
\label{gl_k}
 {\cal J}_i^- &=& \frac{\pa}{\pa x_i},\qquad \quad
i=1,2\ldots k \ , \non \\
 {{\cal J}_{ij}}^0 &=&
x_i \frac{\pa}{\pa x_j}, \qquad i,j=1,2\ldots k \ ,
 \non \\
{\cal J}^0 &=& \sum_{i=1}^{d} x_i\frac{\pa}{\pa x_i}-n\, ,
 \\
 {\cal J}_i^+ &=& x_i {\cal J}^0 =
x_i\, \left( \sum_{j=1}^{d} x_j\frac{\pa}{\pa x_j}-n \right),
\quad i=1,2\ldots k \ . \non
\end{eqnarray}
where $n$ is any real or complex number. It is easy to check that
$J$'s obey the commutation relations of the algebra $gl_{k+1}$.
Furthermore, if $n=0,1,2\ldots$, then the {\it finite-dimensional}
irreps appear with representation spaces
\[
{\cal P}_n^{(k)}\ =\ \langle {x_{\scriptscriptstyle 1}}^{p_1}
{x_{\scriptscriptstyle 2}}^{p_2}\ldots
 {x_{\scriptscriptstyle  d}}^{p_{k}}
 \vert \ 0 \le \Sigma p_i \le n \rangle
\]

{\bf Remark.} The flag  ${\cal P}^{(k)}$ is made out of
irreducible finite-dimensional representation spaces of the
algebra $gl_{k+1}$ taken in realization (\ref{gl_k}).

There exist other flags associated with irreducible
finite-dimensional representation spaces of the Lie algebras of
differential (difference) operators.

{\it Definition:}
\begin{quote}
The operator $h$ is called {\em algebraic}, if it preserves a flag
of polynomials. It has the form $\sum \mbox{Pol}_n \cdot \pa^n$.
This form is called {\it algebraic}.
\end{quote}

It is evident that the existence of an algebraic form does {\it
not} guarantee that the operator preserves the flag of
polynomials.

{\bf Theorem 1.}
\begin{quote}{\it
Linear differential operator $h$ preserves the flag  ${\cal
P}^{(k)}$ iff $h= P({\cal J}(b\subset gl_{k+1})) $, where $P$ is a
polynomial in generators of the maximal affine subalgebra $b$ of
the algebra $gl_{k+1}$, being taken in realization (\ref{gl_k}). }
\end{quote}

In particular, {\it exactly-solvable} second order differential
operator $h$ (preserving the flag  ${\cal P}^{(k)}$) has the form
\begin{equation}
\label{hyper-gl_k}
 h = P_2^{(ij)}(x) \pa_{i}\pa_{j} + P_1^{(i)}(x) \pa_i
\end{equation}
where $P_2^{(ij)}(x)$ and $ P_1^{(i)}(x)$ are the 2nd and 1st
degree polynomials in coordinates $x$'s with arbitrary
coefficients. It is the standard multidimensional {\it
hypergeometrical} operator. It was shown in \cite{Ruhl:1995,
Brink:1997} that the Calogero and Sutherland models (equivalently,
the $A_k$-rational and trigonometric models), and the
${BC}_k$-rational and trigonometric models are of the
hypergeometrical type (\ref{hyper-gl_k}). In order to get these
models in the form (\ref{hyper-gl_k}) in the Hamiltonian the
ground state eigenfunction should be taken as a gauge factor and
then Weyl-invariant (trigonometric) polynomials have to be taken
as new variables.

Another natural hypergeometrical operator appears in connection to
the flag ${\cal P}^{(1,2)}$.  Let us consider the algebra
$g^{(2)}$ generated by
\begin{eqnarray}
\label{g2}
 L^1 = \pa_1    \quad   ,   &   L^2 = x_1\pa_1 -n/3 \ , &\non \\
 L^3 = x_2\pa_2 -n/6 \quad  ,   &
 L^4 = x_1^2\pa_1 + 2x_1 x_2\,\pa_2 -n x_1\ , &\non \\
 L^5 = \pa_2 \quad  ,         &   L^6 = x_1\pa_2\ , &\non \\
 L^7 = x_1^2\pa_2 \quad , &  T=x_2\pa^2_{11}\ ,
\end{eqnarray}
(see \cite{Rosenbaum:1998}), where $n$ is any real or complex
number. If $n=0,1,2, \ldots $, the algebra $g^{(2)}$ has
finite-dimensional irreps with representation spaces given by
${\cal P}_n^{(1,2)}$. These representation spaces form the flag
${\cal P}^{(1,2)}$. It is rather obvious that except for the
operator $L^4$ the remaining operators in (\ref{g2}) has the space
${\cal P}_n^{(1,2)}$ as common invariant subspace for any $n$.
Therefore the second degree differential operator preserving the
flag ${\cal P}^{(1,2)}$ is constructed by taking the second degree
polynomial in $L_{1,2,3,5,6}$ plus the generator $T$. Its explicit
form is
\[
 h = (a^{(11)}_1 x_1^2 + a^{(11)}_2 x_1 + a^{(11)}_3 x_2 + a^{(11)}_4)
 \pa_{1}\pa_{1}
\]
\[
 +
 (a^{(12)}_1 x_1^3 + a^{(12)}_2 x_1^2  + a^{(12)}_3 x_1 x_2 +
 a^{(12)}_1 x_4 + a^{(12)}_5 x_2 + a^{(12)}_6) \pa_{1}\pa_{2} +
\]
\[
 (a^{(22)}_1 x_1^4 + a^{(22)}_2 x_1^3 + a^{(22)}_3 x_1^2 x_2 +
 a^{(22)}_4 x_1^2 + a^{(22)}_5 x_2^2
 + a^{(22)}_6 x_1 x_2 + a^{(22)}_7 x_1 + a^{(22)}_8 x_2  + a^{(22)}_9)
 \pa_{2}\pa_{2}  +
\]
\begin{equation}
\label{hyper-gl_2}
  (b^{(1)}_1 x_1 + b^{(1)}_2) \pa_1  +
 (b^{(2)}_1 x_1^2 + b^{(2)}_2 x_1 + b^{(2)}_3 x_2 + b^{(2)}_4)
 \pa_2\ .
\end{equation}
It was shown in \cite{Rosenbaum:1998} that the algebraic form of
both rational and trigonometric $G_2$ models are a particular type
of (\ref{hyper-gl_2}). It can be easily checked that the flag
${\cal P}^{(1,2)}$ is minimal for these models.

There exist other hypergeometrical operators related with flags of
polynomials other than ${\cal P}^{(k)}$ or ${\cal P}^{(1,2)}$. In
Table I we enlist the characteristic vectors of minimal flags of
the Olshanetsky-Perelomov Hamiltonians.

\section{\bf Perturbation theory}

Existence of algebraic forms of exactly-solvable operators allows
the construction of an {\it algebraic} perturbation theory, where
the procedure of finding corrections is {\bf linear algebra
procedure}.

Consider the spectral problem,
\begin{equation}
\label{pt.1}
 (h_0 + \la h_1)\phi = E \phi \ ,
\end{equation}
where $\la$ is a formal parameter, the solution of that we look
for in perturbation theory,
\begin{equation}
\label{pt.2}
 \phi = \sum \la^k \phi_k \ ,\ E = \sum \la^k E_k \ .
\end{equation}
Then the following theorem holds:

{\bf Theorem 2} \cite{Turbiner:2002}
\begin{quote}{\it
 Let $h_0$ be an exactly-solvable operator with flag ${\cal V}$.
 Let the perturbation $h_1 = {v}_1 {\bf 1} $ is such
 that (i) ${v}_1$ is an element of a finite-dimensional space
 ${\cal V}_n$ from the flag  and (ii) we look for $\phi \in {\cal V}$.
 Then the perturbation theory is algebraic:
 $\exists p(k)$ such that $k$th correction $\phi_k \in {\cal V}_{p(k)}$
 and hence can be found algebraically.
 }
 \end{quote}

Equation to solve to find $k$th correction is:
\[
(h_0 - E_0) \phi_k = \sum_{i=1}^{k} E_i \phi_{k-i} - v_1
\phi_{k-1}
\]

As an immediate consequence of Theorem it can be proven the
following. If $h_0$ depends on some parameters holomorphically
that the coefficients in $\phi_k$ and $E_k$ are rational functions
parameters. In the case of absence of parameters in $h_0$ the
coefficients in $\phi_k$ and $E_k$ are rational numbers.

This form of the perturbation theory is characterized by a certain
self-similarity property -- a calculation of $k$th correction is
similar to the calculation of the first correction but with a
modified perturbation potential $$v_k = -(\sum_{i=1}^{k-1} E_i
\phi_{k-i} - v_1 \phi_{k-1})/\phi_0\ .$$

\section{\bf Algebraic forms of the Olshanetsky-Perelomov Hamiltonians}

As was indicated in the previous Section the existence of the
algebraic forms of exactly-solvable problems allows for a certain
classes of perturbations to construct perturbation theory
algebraically. Thus, the exactly-solvable Olshanetsky-Perelomov
Hamiltonians for which there exist algebraic forms can be possible
unperturbed problems for developing such a perturbation theory. We
present those algebraic forms below.

In order to find the algebraic forms of the Olshanetsky-Perelomov
Hamiltonians \cite{Olshanetsky:1983} the following strategy is
used \cite{Ruhl:1995, Brink:1997, Rosenbaum:1998, Boreskov:2001},
\begin{itemize}
    \item Gauging away ground state eigenfunction, $(\Psi_{0})^{-1}\,
 {\cal H} \Psi_{0}$,
    \item Olshanetsky-Perelomov Hamiltonians possess different
symmetries (permutations, translation-invariance, reflections,
periodicity etc). These symmetries correspond to the Weyl group
acting on the root space. By coding these symmetries to new
coordinates we find 'premature' operators to these Hamiltonians,
 \begin{itemize}
    \item Rational case -- Weyl-invariant variables:
\[
 t_{a}^{(\Om)}(x) = \sum_{\al\in\Om} (\al,x)^{a}\ ,
\]
where $a$'s are the {\em degrees} of the Weyl group $W$ and $\al$
is an orbit. They are defined ambiguously depending on the orbit
taken, but they {\bf always lead to algebraic forms},
   \item Trigonometric case -- trigonometric (periodic)
Weyl-invariant variables,
\[
 \ta_{a}^{(\Om)}(x) = \sum_{\al\in\Om} {\sin}^{a}(\al,x)\ ,
\]
Not always algebraic forms appear in this case, only for very
special combination of $\ta_{a}^{(\Om)}$ it emerges.
 \end{itemize}
\end{itemize}

\bigskip

\subsection{\it HISTORY}

Rational Case:

\begin{itemize}

\item Flat-space metric $g^{\mu \nu}$ being written in
$t_{a}^{(\Om)}$-coordinates has polynomial matrix elements
(V.I.Arnold, '76 \cite{Arnold:1976})

\item Laplacian and $A_n$-extended Laplacian (it means
$A_n$-rational model or Calodero Model) in $t_{a}^{(\Om)}$ --
invariants of $A_n$ root space -- have an algebraic form and are
written in generators of $gl_n$-algebra of 1st order differential
operators (\ref{gl_k}) (W.Ruhl, A.T., '95 \cite{Ruhl:1995})

\item Laplacian and $(B,C,BC)_n$-extended Laplacian (in other
words, $(B,C,BC)_n$-rational model) in $t_{a}^{(\Om)}$ --
invariants of $(B,C,BC)_n$ root space -- have an algebraic form
and are written in generators of $gl_n$-algebra of 1st order
differential operators (\ref{gl_k}) (L.Brink, A.T., N.Wyllard, '98
\cite{Brink:1997})

\item Similar for $G_2$ and $F_4$ rational models but with hidden
algebras $g^{(2)}$ and $f^{(4)}$, resp. (M.Rosenbaum et al, '98
\cite{Rosenbaum:1998}, W.Ruhl et al, '00 \cite{Ruehl:1998} and
K.G.Boreskov et al, '01 \cite{Boreskov:2001}) as well as
$E_{6,7,8}$ \cite{Boreskov:2003}.
\end{itemize}

Trigonometric Case:
\begin{itemize}

\item Laplacian and $A_n$-extended Laplacian ($A_n$-trigonometric
model or Sutherland Model) in $\ta_{a}^{(\Om)}$ -- trig.invariants
of $A_n$ root space -- have an algebraic form and are written in
generators of $gl_n$-algebra of 1st order differential operators
(\ref{gl_k}) (W.Ruhl \& A.T., '95 \cite{Ruhl:1995})

\item Laplacian and $(B,C,BC)_n$-extended Laplacian
($(B,C,BC)_n$-trigonometric model) in $\ta_{a}^{(\Om)}$ --
trig.invariants of $(B,C,BC)_n$ root space -- have an algebraic
form and are written in generators of $gl_n$-algebra of 1st order
differential operators (\ref{gl_k}) (L.Brink, A.T., N.Wyllard, '98
\cite{Brink:1997})

\item Similar for $G_2$ and $F_4$ trigonometric models but with
hidden algebras $g^{(2)}$ and $f^{(4)}$, resp. (M.Rosenbaum et al,
'98 \cite{Rosenbaum:1998}, W.Ruhl et al, '00 \cite{Ruehl:1998} and
K.G.Boreskov et al, '01 \cite{Boreskov:2001})
\end{itemize}

\subsection{\it Algebraic forms}

The algebraic forms of the Olshanetsky-Perelomov Hamiltonians are
given by the following differential operator
\begin{equation}
\label{h}
 h= {\cal A}_{ij}(\ta) \frac{\pa^2}{\pa
 {\ta_i} \pa {\ta_j} } + {\cal B}_i(\ta) \frac{\pa}{\pa \ta_i}
\end{equation}
with polynomial coefficient functions:

{\bf 1.} $A_{n}$ rational case (Calogero model)

\begin{eqnarray}
\label{cal}
 \hspace{-10pt}{\cal A}_{ij}\hspace{-9pt} &=& \hspace{-5pt}
\frac{(n-i+1)j}{n+1}\,\tau_{i}\,\tau_{j} + \hspace{-20pt}
\sum_{{l\geq}{\max (1,j-i)}} \hspace{-15pt}
(j-i-2l)\,\tau_{i+l}\,\tau_{j-l}\ ,  \non
\\[10pt]
 {\cal B}_i \hspace{-9pt}  &=& \hspace{-5pt}
 - \frac{1}{n+1}(1+\nu+\nu n){(n-i+2)(n-i+1)}\,\tau_{i-1} +2\om \,(i+1)\,
\tau_i\ ,
\end{eqnarray}
where $i,j=1,2,\ldots n$.

{\bf 2.} $A_{n}$ trigonometric case (Sutherland model)

\begin{eqnarray}
{\cal A}_{ij}\hspace{-10pt} &=& \hspace{-5pt}
\frac{(n+1-i)\,j}{n+1}\,\ta_{i}\,\ta_{j} + \hspace{-10pt}
\sum_{{l\geq}{\max (1,j-i)}} \hspace{-15pt}
(j-i-2l)\,\ta_{i+l}\,\ta_{j-l}\ ,
  \non \\ {\cal B}_i
\hspace{-10pt}  &=& \hspace{-5pt}
  \frac{1}{n+1}(1+\nu + \nu n)\,i\,(n+1-i)\,\ta_{i} \ ,
\end{eqnarray}
where $i,j=1,2,\ldots n$.

{\bf 3.} $BC_n$ rational model

\begin{eqnarray}
{\cal A}_{ij} &=& 4\, \sum_{l\ge 0} (2l+1+j-i)\,
\ta_{i-l-1}\,\ta_{j+l}\ ,
 \non \\[10pt]
{\cal B}_i &=& 2\, \left[ 1+\nu_2 + 2\nu(n-i)\right] (n-i+1)\,
\ta_{i-1} -4\,\om\,i\,\ta_i \ ,
\end{eqnarray}
where $i,j=1,2,\ldots n$.

{\bf 4.} $BC_n$ trigonometric model

\begin{eqnarray}
\hspace{-10pt}{\cal A}_{ij} \hspace{-10pt}&=&\hspace{-10pt}
n\,{\ta}_{i-1}\,\ta_{j-1} - \hspace{-5pt} \, \sum_{l\ge 0} \Big[
  (i-l)   \,{\ta}_{i-l}  \,\ta_{j+l}
+ (l+j-1) \,\ta_{i-l-1}\,\ta_{j+l-1}
 \non\\
&& - (i-2-l) \,\ta_{i-2-l}\,\ta_{j+l} - (l+j+1)
\,\ta_{i-l-1}\,\ta_{j+l+1} \Big]\ ,
 \non\\[5pt]
{\cal B}_i  \hspace{-10pt}&=&\hspace{-10pt}
\frac{\nu_3}{2}(i-n-1)\,{\ta}_{i-1} - \Big[1+\nu_2 +
\frac{\nu_3}{2}+\nu(2n-i-1)\,\Big]i\,\ta_{i}
 \non\\
&&  \hspace{110pt} - \nu(n-i+1)(n-i+2)\ta_{i-2} \ ,
\end{eqnarray}
where $i,j=1,2,\ldots n$.

{\bf 5.} $G_2$ rational model

\[
{\cal A}_{11} = 2\ta_1\ ,\ {\cal A}_{12}=  12\ta_2\ ,\ {\cal
A}_{22} = -\frac{8}{3}\ta_1^2\ta_2\ ,
\]
\begin{equation}
 \label{g2r}
   {\cal B}_1=  \frac{4}{3}\om\ta_1+2(1+3\mu+3\nu)\ ,\
   {\cal B}_2= 4\om\ta_2-\frac{4}{3}(1+2\nu)\ta_1^2\ .
\end{equation}

{\bf 6.} $G_2$ trigonometric model

\[
{\cal A}_{11} =
(2\ta_1+\frac{\al^2}{2}\ta_1^2-\frac{\al^4}{24}\ta_2)\ ,\ {\cal
A}_{12}=  (12+\frac{8\al^2}{3}\ta_1)\ta_2\ ,\ {\cal A}_{22}=
-(\frac{8}{3}\ta_1^2\ta_2-2\al^2\ta_2^2)\ ,
\]
\begin{equation}
 \label{g2t}
   {\cal B}_1=  2(1+3\mu +3\nu) + \frac{2}{3}(1 + 3\mu +
       4 \nu)\al^2\ta_1 \ ,\
   {\cal B}_2= - \frac{4}{3}(1+2\nu)\ta_1^2 + [\frac{7}{3}
 +4(\mu +\nu )]\al^2\ta_2\ .
\end{equation}

Algebraic forms for the $F_4$ rational and trigonometric models as
well as $E_6,7,8$ rational models can be found in
\cite{Boreskov:2001} and \cite{Boreskov:2003}, respectively.

All rational models possess a remarkable property - each of them
is characterized by appearance of infinite family of eigenstates
depending on single variable, which is the second Weyl invariant
$t_2$ (see definition above) \footnote{The invariant $t_2$ has
unique feature -- it is the same for all Weyl groups.}. This
family always includes the ground state. Finding these states is
reduced to solving a simple eigenvalue problem for one-dimensional
differential operator, which can be written in the form
\[
 h_{r}\ =\ -2 t_2\pa^2_{ t_2 t_2} +
 (4\om t_2 - 1 - 2a_r)\pa_{t_2}\ ,
\]
where the parameter $a_r$ depends on the model studied. It can be
easily demonstrated that the operator $h_{r}$ is reduced to the
Hamiltonian of the $A_1$ rational model
\[
{\cal H}_{\rm A_1} = - \frac{1}{2} \sum_{i=1}^{2} \frac{\pa^2}{\pa
{x_i}^2} + \frac{\om^2}{2} (x_1-x_2)^2 + \frac{g}{({x_1}-x_2)^2}
\]
where the coupling constant $g=a_r(1-a_r)$.

\section{\bf Perturbation theory (examples)}

In this Section we present several seemingly interesting examples
of perturbative calculations for the ground state for different
perturbations. Other examples can be found at
Ref.\cite{Turbiner:2002}.

{\bf 1.} $G_2$ rational model (see \ref{h} with coefficients
(\ref{g2r})).

Let us take $\ta_2$ as a perturbation which in Cartesian
coordinates looks like
\[
v_1^{(r)} = \ta_2 = (x_1-x_2)^2 (x_2-x_3)^2(x_1-x_3)^2
\]
The ground state is
\[
 \phi_0=1\ ,\ \ep_0=0\ .
\]
After simple calculations we get
\[
\phi_1\ =\ -\frac{(1+2\nu)}{8\om^2} \ta_1^2 - \frac{1}{4\om} \ta_2
+ \frac{3}{8\om^3} (1+2\nu)(2+3\mu+3\nu)\ta_1\ ,
\]
\begin{equation}
\ep_1 = \frac{3}{4\om^3} (1+2\nu)(1+3\mu+3\nu)(2+3\mu+3\nu)\ .
\end{equation}

It is not very complicated to calculate next several corrections.

{\bf 2.} $G_2$ trigonometric model (see \ref{h} with coefficients
(\ref{g2t})).

{\it (i)}. Take $\ta_1$ as a perturbation, which in Cartesian
coordinates is
\[
v_1^{(t)} = \ta_1\  =
\frac{-2}{\al^2}\big[\sin^2\frac{\al}{2}(x_1-x_2)+\sin^2\frac{\al}{2}(
x_2-x_3)+\sin^2\frac{\al}{2}(x_3-x_1)\big] \
\]
The first correction to the ground state is given by
\[
\phi_1\ =\ - \frac{3}{2(1+3\mu+4\nu)\al^2}\ta_1\ ,
\]
 and
\begin{equation}
\label{g2t1}
 \ep_1=-\frac{3(1+3\mu+3\nu)}{(1+3\mu+4\nu) \al^2}\ .
\end{equation}
It is worth noting that $\phi_1$ depends on the single variable
$\ta_1$. However, it will not be the case for higher corrections
they begin to depend on both $\ta_{1,2}$. The expression
(\ref{g2t1}) has quite amusing property. If $\nu=0$ that
corresponds to the three-body Sutherland model, $\ep_1$ does not
depend on $\mu$. Of course, it does not hold for higher
corrections but leads to rather non-trivial property of some
correlation function (see below).

{\it (ii)}. Take $\ta_2$ as a perturbation, which in Cartesian
coordinates has the form
\[
v_1^{(t)} = \ta_2  =  \frac{16}{\al^6} \biggl[
\sin\al(x_1-x_2)+\sin\al( x_2-x_3)+\sin\al(x_3-x_1)\biggr]^2 \
\]
In the limit $\al \rar 0$, the perturbation $v_1^{(t)}$ becomes
$v_1^{(r)}$. The first correction to the ground state is given by
\[
\phi_1\ =\ -\frac{1}{N}\bigg[\frac{12(1+2\nu)}{\al^2} \ta_1^2 +
3(7+12\mu+16\nu) \ta_2 - 72
\frac{(2+3\mu+3\nu)(1+2\nu)}{(1+3\mu+4\nu)\al^4}\ta_1\bigg]
\]
 and
\begin{equation}
\ep_1=\frac{ 144(1+2\nu)(1+3\mu+3\nu)(2+3\mu+3\nu)}{(1+3\mu+4\nu)
N\al^4}\ ,
\end{equation}
where $N=[7(7+12\mu+16\nu) - (1-84\mu-82\nu
-144\mu^2-336\mu\nu-192\nu^2)\al^2]$. It is not very complicated
to calculate next several corrections.

{\bf 3.} $A_n$ rational model (in other words, the $(n+1)$-body
Calogero problem, (see \ref{h} with coefficients (\ref{cal}))).
Let us take $\ta_3$ as a perturbation. In Cartesian coordinates it
looks like
\[
v_1^{(r)} = \ta_3 = \si_{4}(y) = \sum_{i_{1},i_{2},i_{3},i_{4}}
y_{i_{1}}y_{i_{2}}y_{i_{3}}y_{i_{4}}\ ,
\]
where $\si_{4}(y)$ is the fourth order elementary symmetric
polynomial of the Perelomov coordinates as arguments,
\[
Y=\sum x_i\ ,\ y_i=x_i - \frac{1}{n+1} Y\ ,\ i=1,\ldots , (n+1) \
.
\]
Easy calculation gives quite simple answer for the first
correction

\[
-\phi_1 = \frac{1}{8\om}\,\ta_3 + \frac{[1+\nu
(n+1)](n-1)(n-2)}{32(n+1)\om^2} \,\ta_1 \ ,
\]
\begin{equation}
\label{cal1}
 \ep_1= \frac{[1+\nu (n+1)]^2 n(n-1)(n-2)}{32(n+1)\om^2}\ .
\end{equation}
It can be shown that for non-physical value $n=1$ (somehow
`one-body' problem) the energy correction $\ep_1$ as well as all
other corrections vanish. A meaning of this result is not clear so
far.

\section{\bf Generalized Dyson-Mehta integrals}

Let us take one of crystallographic root spaces $R (A_n, B_n, C_n,
D_n, BC_n, G_2, F_4, E_{6,7,8})$. Suppose $P_c (t)$ is a
Weyl-invariant polynomial of finite order in $R$ and $D_c =
\mbox{Weyl chamber}$
\[
 E_{\nu} (P_c) = \int_{D_c} P_c(t) (\mbox{Weyl det})^{\nu} \exp
({-\om t_2}) dt
\]
$t_2$ is 2nd invariant. Similarly one can introduce $Pt (\ta)$ as
the Weyl-invariant trigonometric polynomial of finite order in $R$
and $D_t = \mbox{Weyl alcove}$,
\[
 T_{\nu} (Pt) = \int_{D_t} Pt (\ta) (\mbox{Trig. Weyl det})^{\nu}
 d\ta
\]
$E_{\nu}, T_{\nu}$ are related to the Selberg integrals and we
call them {\it generalized Dyson-Mehta integrals}. If $P_c(t)\
(P_t(\ta))=1$, they become the Dyson-Mehta integrals (for
discussion see \cite{Olshanetsky:1983}).

{\bf Theorem 3.}

(i) $\frac{E_{\nu} (P_c)}{E_{\nu} (P_c=1)}$ is rational function
in $\nu, \om$ with rational coefficients,

(ii) $\frac{T_{\nu} (P_t)}{T_{\nu} (P_t=1)}$ is rational function
in $\nu$ with rational coefficients.

\bigskip

{\bf Proof.}
\begin{itemize}
    \item Weyl determinant defines the ground state eigenfunction of
    $R$-inspired rational (trigonometric) model,
    \item $\exists$ a set of Weyl invariants which being taken as
    variables lead to algebraic form of the Hamiltonian, they are
    generators of algebra of Weyl-invariant polynomials,
    \item Following the Theorem 2 the perturbation theory for any
    $P_c(t)\ (P_t(\ta))$ is algebraic. Hence the first energy
    correction to the ground state is found by pure algebraic means.
    Since parameters enter to the algebraic forms of the
    unperturbed operators linearly or at most quadratically, it is
    evident that the answer can contain not more than rational
    function of parameters. {\it Q.E.D.}
\end{itemize}

It is worth to remind that for $A_N$,
\[
(\mbox{Weyl det})\ \doteq \ \prod_{i<j}^N (x_i-x_j)
\]
\[
(\mbox{Trig. Weyl det}) \ \doteq \ \prod_{i<j}^N \sin (x_i-x_j)
\]
are anti-symmetric invariants.

A constructive way to realize Theorem 3 is to perturb one of the
Olshanetsky-Perelomov Hamiltonians with a perturbation, which
meets the conditions of the Theorem. Then make two calculations of
the first energy correction: (i) one is based on the algebraic
perturbation theory and (ii) another is based on the
Rayleigh-Schroedinger Perturbation Theory (RSPT). Now let us
consider the examples presented in the previous section.\\[10pt]

\centerline{\bf EXAMPLES}

Case {\bf 2(i).}
\[
\frac{ \displaystyle \int_{D_t(G_2)}
 \big[\sin^2\frac{\al}{2}(x_1-x_2)+\sin^2\frac{\al}{2}(
x_2-x_3)+\sin^2\frac{\al}{2}(x_3-x_1)\big]
 W^2 (x_1,x_2,x_3)\, d^3 x
 }
 {
 \displaystyle \int_{D_t(G_2)} \, W^2 (x_1,x_2,x_3)\, d^3 x}\ =
\]
\[
   =\ \frac{3(1+3\mu+3\nu)}{2(1+3\mu+4\nu)} \ ,
\]
where $D_t(G_2)$ is the $G_2$ Weyl alcove and $$W (x_1,x_2,x_3)=
\prod_{i<j}^3 \sin^{\nu}\frac{\al}{2}(x_{i}-x_{j})\prod_{i<j}^3
\sin^{\mu}\frac{\al}{2}(x_{i}-x_{j}-2x_k)\ .$$

Case {\bf 2(ii).}

\[
 \frac{ \displaystyle \int_{D_t(G_2)}
 \biggl[
\sin\al(x_1-x_2)+\sin\al( x_2-x_3)+\sin\al(x_3-x_1)\biggr]^2
 W^2 (x_1,x_2,x_3)\, d^3 x
 }
 {
 \displaystyle \int_{D_t(G_2)} \, W^2 (x_1,x_2,x_3)\, d^3 x}\ =
\]
\[
 =\ \frac{9
\al^2(1+2\nu)(1+3\mu+3\nu)(2+3\mu+3\nu)}{(1+3\mu+4\nu)N}\ ,
\]

Case {\bf 3.}

\[
\frac{\displaystyle \int_{D_c(A_{n-1})}
(\sum_{\{i_{1},i_{2},i_{3},i_{4}\}}
y_{i_{1}}y_{i_{2}}y_{i_{3}}y_{i_{4}})
\prod_{i<j}|y_{i}-y_{j}|^{2\nu} e^{-\om\sum{y_i^{2}}} d^{n-1}y
}{\displaystyle \int_{D_c(A_{n-1})}
\prod_{i<j}|y_{i}-y_{j}|^{2\nu} e^{-\om\sum{y_i^{2}}} d^{n-1}y}\ =
\]
\[
 =\ \frac{1}{32\om^2}\frac{(1 + n\nu)^2
(n-1)(n-2)(n-3)}{n}\ ,
\]
where $D_c(A_{n-1})$ is the $A_{n-1}$ Weyl chamber.


\begingroup\raggedright
\endgroup

\end{document}